\documentclass[twocolumn,showpacs,superscriptaddress,amssymb,10pt,pra,floatfix]{revtex4}

\usepackage{graphicx}
\usepackage{dcolumn}
\usepackage{bm}
\usepackage{epsfig}
\usepackage{color}
\usepackage{longtable}

\def\ketm#1{  \left\vert  #1   \right\rangle   }

\def\sprm#1#2{  \left\langle #1 \left\vert \right. #2 \right\rangle   }

\def\mem#1#2#3{  \left\langle #1 \left\vert  #2 \right\vert #3 \right\rangle   }

\def\rmem#1#2#3{  \left\langle #1 \left\vert \left\vert  #2
                  \right\vert \right\vert #3 \right\rangle   }
\begin{document}

\preprint{}
%
%
%
%
\title{Modification of multipole transitions by twisted light}
%
%
%
%

\author{S.~A.-L.~Schulz}
\email[]{sabrina.schulz@ptb.de}

\affiliation{Physikalisch--Technische Bundesanstalt, D--38116 Braunschweig, Germany}
\affiliation{Technische Universit\"at Braunschweig, D--38106 Braunschweig, Germany}

\author{S.~Fritzsche}
\affiliation{Helmholtz--Institut Jena, D--07743 Jena, Germany}
\affiliation{Theoretisch--Physikalisches Institut, Friedrich--Schiller--Universit\"at Jena, D--07743 Jena, Germany}

\author{R.~A.~M\"uller}
\affiliation{Physikalisch--Technische Bundesanstalt, D--38116 Braunschweig, Germany}
\affiliation{Technische Universit\"at Braunschweig, D--38106 Braunschweig, Germany}

\author{A.~Surzhykov}
\affiliation{Physikalisch--Technische Bundesanstalt, D--38116 Braunschweig, Germany}
\affiliation{Technische Universit\"at Braunschweig, D--38106 Braunschweig, Germany}

\date{\today \\[0.3cm]}

%
%
%
%

\begin{abstract}
A theoretical analysis is presented for the excitation of single many--electron atoms and ions by twisted (or vortex) light. Special emphasis is put on excitations that can proceed via several electric and magnetic multipole channels. We argue that the relative strength of these multipoles is very sensitive to the topological charge and kinematic parameters of the incident light and can be strongly modified with respect to the plane--wave case. Most remarkably, the modification of multipole transitions by twisted radiation can be described by means of a \textit{geometrical} factor. This factor is  independent of the shell structure of a particular target atom and just reflects the properties of the light beam as well as the position of an atom with respect to the vortex axis. An analytical expression for the geometrical factor is derived for Bessel photons and for a realistic experimental situation in which the position of an atom is not well determined. To illustrate the use of the geometrical factor for the analysis of (future) measurements, detailed calculations are presented for the $3s3p \, ^3P_1 \to 3s 3p \, ^1P_1$ excitation in neutral Mg.   
\end{abstract}

\maketitle

%
%
\section{Introduction}

In recent years, a considerable interest has been given to the application of twisted (or vortex) light beams in atomic physics. These beams, with their helical phase fronts, non--uniform intensity profiles and complex internal flow patterns \cite{BeB11}, can significantly modify the fundamental photo--induced atomic processes. A number of studies have been performed, for example, to investigate how the special properties of twisted light affect atomic photo--ionization \cite{MaH13, SuS16} as well as Rayleigh, Raman and Compton scattering \cite{PeV18,JeS11,Vie16}. 

The photo--excitation of single trapped atoms and ions is also in the focus of intense experimental and theoretical research. It was shown, in particular, that the use of vortex light can lead to a significant reduction of the light shift of bound--state transitions \cite{ScS16}. This opens new opportunities for the high--precision spectroscopy of electric dipole--forbidden clock transitions. Moreover, the selection rules, that relate the angular momenta $J$ and their projections $M$ in the excitation $\ketm{J_i M_i} + \gamma \to \ketm{J_f M_f}$, can be remarkably modified if $\gamma$ is a twisted and not a plane--wave photon \cite{AfC13,AfC18}. Such a modification of selection rules leads to the fact that the relative strength of allowed \textit{multipole components} of a particular bound--state transition can be changed as well. A first step in the theoretical analysis of the operation of multipole components by twisted light has been done recently by Afanasev and co--workers \cite{AfC18b}. It was shown, in particular, that the contributions of magnetic dipole (M1) and electric quadrupole (E2) channels to the $^2P_{1/2} \to ^2D_{3/2}$ transition in boron--like ions strongly depend on the position of the target ion within incident twisted beam. For the preparation and analysis of forthcoming photo--excitation experiments, however, a general formalism has to be laid out for the interaction of vortex light with an \textit{arbitrary} many--electron atom,  independent of its shell structure. This formalism has to account, moreover, for the realistic experimental situation in which the position of the target atom (or ion) within the photon wave front is not well determined. 

In this contribution, we present a theoretical study of the excitation of a single many--electron atom by twisted light. Our work deals especially with the scenario in which the excitation proceeds via \textit{several} electric and magnetic multipole channels. To start a discussion of whether and how the strength of these multipoles is affected by twisted light, we shall remind first the well--known plane--wave results. In Sec.~\ref{subsec:theory_plane_wave}, therefore, we briefly consider the evaluation of the plane--wave matrix elements and transition rates. Their counterparts for twisted beams are derived then in Sec.~\ref{subsec:theory_twisted_wave}. We show, in particular, that for transitions between states with unresolved magnetic sublevel structure the multipole rates are added with \textit{no} interference term. Therefore, it is natural to quantify the relative contribution of the  multipole components of twisted light to a particular transition by the ratio of their rates. In Sec.~\ref{subsubsec:relative_contribution_twisted} we show that this ratio can be written as a product of two terms. While the first term is just a ratio of the usual plane--wave multipole transition rates, the second is given by the so--called geometrical factor. This factor is  independent of the shell structure of a particular atom but is sensitive to the topological charge and kinematic properties of incident Bessel light. Detailed calculations of the geometrical factor are presented in Sec.~\ref{subsec:results_geometrical_factors} for dipole, quadrupole and octupole transitions. Special emphasis in these calculations is placed upon the uncertainty in localizing the atom within the light front. In particular, we analyze the $3s3p \, ^3P_1 \to 3s 3p \, ^1P_1$ excitation of a neutral Mg atom. In Sec.~\ref{subsec:results_Mg_transition} we show that while for the incident plane wave both magnetic dipole (M1) and electric quadrupole (E2) components contribute almost equally to this line, the relative strength of the E2 term can be enhanced by almost an order of magnitude if a Bessel beam with topological charge \textit{two} is used for the excitation. The summary of these results and outlook are given finally in Sec.~\ref{sec:summary}. 

Hartree atomic units ($\hbar = e = m_e = 1$) are used throughout the paper unless stated otherwise.

%
%
\section{Theoretical background}
\label{sec:theory}

\subsection{Excitation by plane--wave photons}
\label{subsec:theory_plane_wave}

Since the excitation of atoms by plane--wave radiation has been studied in a large number of works, here we just recall the basic formulas needed for the further analysis. We start our discussion from the transition amplitude:
\begin{equation}
	\label{eq:transition_amplitude_plane_wave}
	M_{fi}^{\rm (pl)} = \mem{\alpha_f J_f M_f}{\sum\limits_{q} {\bm \alpha}_q \, {\bm A}^{\rm (pl)}({\bm r}_q)}{\alpha_i J_i M_i} \, ,
\end{equation}
which is needed to calculate both the excitation rate and all properties of the final (excited) state. This amplitude is written here for a transition between many--electron states, that are characterized by the total angular momenta $J_{i,f}$ and their projections $M_{i,f}$. Moreover, $\alpha_{i,f}$ denote all additional quantum numbers as required for a unique specification of the atomic states.

The electron--photon interaction operator $\hat {\mathcal R} = \sum_{q} {\bm \alpha}_q \, {\bm A}^{\rm (pl)}({\bm r}_q)$ in the amplitude (\ref{eq:transition_amplitude_plane_wave}) is written as a sum of one--particle operators, where ${\bm \alpha}_q$ denotes the vector of Dirac matrices for the $q^{th}$ particle and ${\bm A}^{\rm (pl)}({\bm r}_q)$ is the vector potential of the radiation field. In the Coulomb gauge and for the plane--wave solution this vector potential reads as:
\begin{equation}
	\label{eq:vector_potential_plane_wave}
	{\bm A}^{\rm (pl)}({\bm r}) = {\bm e}_{{\bm k} \lambda} \, {\rm e}^{i {\bm k} {\bm r}} \, ,
\end{equation}
where ${\bm k}$ and ${{\bm e}_{{\bm k} \lambda}}$ are the wave-- and polarization vectors, and $\lambda = \pm 1$ is the helicity of light. For the analysis of the atomic photo--excitation it is very convenient to expand ${\bm A}^{\rm (pl)}({\bm r})$ in terms of its electric and magnetic multipole fields. If the
propagation direction of the light ${\hat {\bm k}} = {\bm k}/k = \left(\theta_k, \phi_k \right)$ does
not coincide with the quantization ($z$--) axis, this expansion is given by:
\begin{eqnarray}
	\label{eq_vector_potential_multipole_decomposition}
	{\bm e}_{{\bm k} \lambda} \, {\rm e}^{i {\bm k} {\bm r}} &=& \sqrt{2 \pi}
	\, \sum\limits_{L M} \sum\limits_{p=0,1} \, i^L \, [L]^{1/2} \, (i \lambda)^p \nonumber \\
	&\times& D^{L}_{M \lambda}(\varphi_k, \theta_k, 0) \, {\bm a}^{(p)}_{LM}({\bm r}) \, .
\end{eqnarray}
Here, $[L] = 2L + 1$, $D^{L}_{M \lambda}$ is the Wigner rotation matrix, and ${\bm a}^{(p)}_{LM}({\bm r})$ refers to magnetic ($p = 0$) and electric ($p = 1$) multipole components. For the sake of brevity, we will not present the explicit form of these components and refer the reader instead to Refs.~\cite{Ros57,Gra74,SuF02}. Here we just mention that ${\bm a}^{(p)}_{LM}({\bm r})$ are constructed as irreducible tensors of rank $L$
which enables further analytical evaluation of the transition amplitude $M_{fi}^{\rm (pl)}$. Indeed, by inserting the expansion (\ref{eq_vector_potential_multipole_decomposition}) into Eq.~(\ref{eq:transition_amplitude_plane_wave}) and making use of the Wigner--Eckart theorem we obtain
\begin{eqnarray}
	\label{eq_plane_wave_matrix_element_multipole_expansion}
	 M_{fi}^{\rm (pl)} &=& \sqrt{2 \pi} \, \sum\limits_{L M} \sum\limits_{p=0,1} \, i^L \, \frac{[L]^{1/2}}{[J_f]^{1/2}} \, (i \lambda)^p \, D^{L}_{M \lambda}( \varphi_k,\theta_k, 0) \nonumber \\[0.2cm]
	&& \hspace*{-0.7cm} \times \sprm{J_i M_i \, L M}{J_f M_f} \, \rmem{\alpha_f J_f}{H_\gamma(pL)}{\alpha_i J_i} \, ,
\end{eqnarray}
where we introduced the short--hand notation
\begin{equation}
	\label{eq_reduced_matrix_element}
	\rmem{\alpha_f J_f}{H_\gamma(pL)}{\alpha_i J_i} = \rmem{\alpha_f J_f}{\sum\limits_q {\bm \alpha}_q \, {\bm a}^{(p)}_{L}({\bm r}_q)}{\alpha_i J_i}
\end{equation}
for the many--electron \textit{reduced} matrix element. This (reduced) matrix element does \textit{not} depend on the projections of the angular momenta of an atom and photon as well as on the particular choice of the quantization axis and the light propagation direction ${\hat {\bm k}}$. It solely reflects the electronic structure of an atom and its coupling to a particular multipole component $(pL)$ of the radiation field. We will see below that these elements $\rmem{\alpha_f J_f}{H_\gamma(pL)}{\alpha_i J_i}$ form the ``building block'' to describe all the properties of the photo--excitation process both for the plane--wave and twisted light.  

With the help of the amplitude $M_{fi}^{\rm (pl)}$ one can calculate now the rate for the $\ketm{\alpha_i J_i} + \gamma \to \ketm{\alpha_f J_f}$ excitation of an atom by the plane--wave light \cite{BrJ83}:
\begin{eqnarray}
	\label{eq_transition_probability_plane_wave}
	W_{fi}^{(pl)} &=& \frac{2 \pi}{[J_i] \alpha^{2}} \sum\limits_{M_i M_f} \left| M_{fi}^{\rm (pl)} \right|^2 \, .
\end{eqnarray}
Here we assumed that the initial atomic state is unpolarized and the magnetic sublevel population of the final state remains unobserved. By inserting Eq.~(\ref{eq_plane_wave_matrix_element_multipole_expansion}) into this expression and making simple angular momentum algebra, we obtain:
\begin{eqnarray}
	\label{eq_transition_probability_plane_wave_2}
	W_{fi}^{\rm (pl)} &=& \sum\limits_{p L} W_{fi}^{\rm (pl)}(p L) ,
\end{eqnarray}
where the partial multipole rate is:
\begin{eqnarray}
	\label{eq_transition_probability_plane_wave_partial}
    W_{fi}^{\rm (pl)}(p L) &=& \frac{(2 \pi)^{2}}{[J_i] \alpha^{2}} \left|\rmem{\alpha_f J_f}{H_\gamma(pL)}{\alpha_i J_i} \right|^2 \, .
\end{eqnarray}
As seen from these formulas, the total (i.e. summed over $M_{i,f}$) photo--excitation rate $W_{fi}^{\rm (pl)}$ is given as the \textit{sum} of rates of individual electric and magnetic multipole transitions. This sum is restricted to the $(pL)$--terms that are allowed by the selection rules for a given choice of initial and final atomic states. In the present work, for example, we consider $\ketm{\alpha_i J_i} + \gamma \to \ketm{\alpha_f J_f}$ transitions which can proceed via two multipole channels.  

\subsection{Excitation by twisted photons}
\label{subsec:theory_twisted_wave}

\subsubsection{Transition amplitudes and rates}

After a brief reminder of basic formulas for the plane--wave radiation, we are ready to explore the excitation of a single trapped atom by twisted light. Again, we start our analysis from the transition amplitude:
\begin{equation}
	\label{eq:transition_amplitude_twisted_wave}
	M_{fi}^{\rm (tw)} = \mem{\alpha_f J_f M_f}{\sum\limits_{q} {\bm \alpha}_q \, {\bm A}^{\rm (tw)}({\bm r}_q)}{\alpha_i J_i M_i} \, ,
\end{equation}
where in contrast to Eq.~(\ref{eq:transition_amplitude_plane_wave}) we have to insert the vector potential ${\bm A}^{\rm (tw)}({\bm r})$ for \textit{twisted} radiation. In the present study we write this potential as
\begin{eqnarray}
	\label{eq_vector_potential_twisted}
	{\bm A}^{\rm (tw)}({\bm r}) &=& {\mathbf A}^{\rm (tw)}_{\varkappa m k_z \lambda}({\bm r}) \nonumber \\
	&& \hspace*{-1cm} =  \int {\bm e}_{{\bm k} \lambda} \, {\rm e}^{i {\bm k} {\bm r}} \, a_{\varkappa m}({\bm k}_\perp) \,
	{\rm e}^{-i {\bm k}_\perp {\bm b}} \, 
	\frac{d^2\bm k_\perp}{(2\pi)^{2}} \, ,
\end{eqnarray}
with the amplitude
\begin{equation}
	\label{eq_vector_potential_twisted_expansion_coefficients}
	a_{\varkappa m}({\bm k}_\perp) =\, {\rm e}^{i m \phi_k} \, \frac{2\pi}{k_\perp} \, \delta(k_\perp-\varkappa) \, .
\end{equation}
These expressions describe so--called \textit{Bessel} photons with  helicity $\lambda$, the longitudinal component $k_z$ of the linear momentum and the projection $m$ of the total angular momentum (TAM) upon the light propagation axis ($z$--axis). Moreover, the Bessel state (\ref{eq_vector_potential_twisted}) is also characterized by the absolute value of the transverse momentum $\left| {\bm k}_\perp \right| = \varkappa$ and, hence, by the photon energy $\omega = k/\alpha =\sqrt{k_z^2 + \varkappa^2}/\alpha$, see Ref.~\cite{JeS11,MaH13,ScF14} for further details. 

The exponential factor ${\rm e}^{-i {\bm k}_\perp {\bm b}}$ in Eq.~(\ref{eq_vector_potential_twisted}) specifies the position of a target atom within the incident beam. One has to introduce this factor since---in contrast to the plane wave case---all properties of twisted beams are position--dependent. For example, the beam intensity profile in the $xy$--plane normal to the propagation ($z$--) axis exhibits a concentric
ring pattern. The impact parameter ${\bm b} = \left( b_x, b_y, 0 \right)$ in Eq.~(\ref{eq_vector_potential_twisted}) is defined with regard to the central point of this ring structure.

Eqs.~(\ref{eq_vector_potential_twisted})--(\ref{eq_vector_potential_twisted_expansion_coefficients}) suggest that a Bessel beam can be interpreted as a coherent superposition of plane waves ${\bm e}_{{\bm k} \lambda} \, {\rm e}^{i {\bm k} {\bm r}}$ whose wave vectors ${\bm k}$ are uniformly distributed upon the surface of a cone with a polar opening angle $\theta_k = \arctan(\varkappa/k_z)$. By making the standard expansion (\ref{eq_plane_wave_matrix_element_multipole_expansion}) of these plane waves and by inserting ${\mathbf A}^{\rm (tw)}_{\varkappa m k_z \lambda}({\bm r})$ into Eq.~(\ref{eq:transition_amplitude_twisted_wave}) we find after simple algebra
\begin{eqnarray}
	\label{eq_transition_amplitude_twisted_wave_expansion}
	M_{fi}^{\rm (tw)} &=& \sum\limits_{LM} \sum\limits_{p=0,1} i^{L+M} \, \frac{[L]^{1/2}}{[J_f]^{1/2}}  \, (i\lambda)^p  \nonumber \\[0.2cm]
	&& \hspace*{-1cm} \times \,  d^L_{M \lambda}(\theta_k) \, J_{m-M}(\varkappa b) \, {\rm e}^{i(m-M)\phi_b} \nonumber \\[0.2cm]
	&& \hspace*{-1cm} \times \sprm{J_i M_i \, L M}{J_f M_f} \, \rmem{\alpha_f J_f}{H_\gamma(pL)}{\alpha_i J_i}
\end{eqnarray}
the amplitude for the  $\ketm{\alpha_i J_i M_i} + \gamma \to \ketm{\alpha_f J_f M_f}$ transition induced by Bessel light. Similar to the plane--wave case (\ref{eq_plane_wave_matrix_element_multipole_expansion}), $M_{fi}^{\rm (tw)}$ is written here as a sum of reduced matrix elements $\rmem{\alpha_f J_f}{H_\gamma(pL)}{\alpha_i J_i}$ weighted by geometric (angular) factors. For Bessel photons these weight factors depend on the opening angle $\theta_k$ and the TAM projection $m$ of the beam as well as on the impact parameter of an atom $b$. These dependences enter Eq.~(\ref{eq_transition_amplitude_twisted_wave_expansion}) through the small Wigner function $d^L_{M \lambda}(\theta_k)$ and the Bessel function $J_{m-M}(b\varkappa)$, respectively. 

By taking the modulus squared of the amplitude (\ref{eq_transition_amplitude_twisted_wave_expansion}) and averaging (summing) it over the magnetic quantum numbers of initial and final states we derive the total rate for the excitation of atoms by Bessel light: 
\begin{eqnarray}
	\label{eq_transition_probability_twisted_wave}
	W_{fi}^{\rm (tw)} &=& \frac{2\pi}{[J_i]\alpha^{2}} \sum\limits_{M_i M_f} \left| M_{fi}^{\rm (tw)} \right|^2 = \sum\limits_{p L} W_{fi}^{\rm (tw)}(p L) \nonumber \\[0.2cm]
	&& \hspace*{-1.5cm} = \sum\limits_{p L} \sum\limits_{M} \left| d^L_{M \lambda}(\theta_k) \, J_{m-M}(b\varkappa) \right|^2 \, W_{fi}^{\rm (pl)}(p L) \, .
\end{eqnarray}
Similar to  $W_{fi}^{\rm (pl)}$, this rate $W_{fi}^{\rm (tw)}$ is a sum of partial rates of allowed multipole transitions $(pL)$. Each partial rate $W_{fi}^{\rm (tw)}(p L)$, moreover, is a product of its plane--wave counterpart (\ref{eq_transition_probability_plane_wave_partial}) and geometrical term that describes the properties of the Bessel beam and the position of the target atom. This implies, therefore, that the relative contributions of the multipole channels $(pL)$ to the total rate of the $\ketm{\alpha_i J_i} + \gamma \to \ketm{\alpha_f J_f}$ transition can be modified by the use of twisted light. 

Before starting a detailed discussion on the modification of multipole transition probabilities by twisted light we note that Eq.~(\ref{eq_transition_probability_twisted_wave}) describes an idealized situation of precise localization of a target atom in the light front. This is not the case in most experiments in which the impact parameter ${\bm b}$ remains undetermined and can at best be described by some mean value. In order to account for such a \textit{delocalization} we assume here that the probability to find an atom at the distance ${\bm b}$ from the beam center is given by:
\begin{eqnarray}
	\label{eq_atom_localization_probability}
	f({\bm b}; {\bm b}_0) &=& \frac{1}{2 \pi \sigma^2} \, {\rm e}^{-\frac{({\bm b} - {\bm b}_0)^2}{2 \sigma^2}}
	\nonumber \\
	&=& \frac{1}{2 \pi \sigma^2} \, {\rm e}^{-\frac{b^2 + b_0^2 - 2 b b_0 \cos\phi_b}{2 \sigma^2}} \, ,
\end{eqnarray}
with ${\bm b}_0$ being the most probable impact parameter and $\sigma$ the width of the distribution. Making use of this expression and Eq.~(\ref{eq_transition_probability_twisted_wave}) we can find the photoexcitation rate for the realistic scenario of a delocalized atom: 
\begin{eqnarray}
	\label{eq_transition_probability_twisted_wave_delocalized}
	W_{fi}^{\rm (tw)}({\bm b}_0, \sigma) && \nonumber \\
	&& \hspace*{-2cm} = \sum\limits_{p L} \sum\limits_{M} \left| d^L_{M \lambda}(\theta_k) \right|^2 \, {\mathcal J}_{m-M}(b_0) \, W_{fi}^{\rm (pl)}(p L) \, .
\end{eqnarray}
Here the function ${\mathcal J}_{m-M}(b_0)$ read as:
\begin{eqnarray}
	\label{eq_cal_J_function}
    {\mathcal J}_{m-M}(b_0) &=& \int f({\bm b}; {\bm b}_0) \, \left| J_{m-M}(\varkappa b)\right|^2 \, {\rm d}{\bm b}  \\
    & & \hspace*{-2cm} =  \frac{1}{\sigma^2} \, \int\limits_{0}^{\infty} {\rm d}b \, b \, I_0\left(\frac{b b_0}{\sigma^2}\right) \, \left|J_{m-M}(\varkappa b) \right|^2 \, {\rm e}^{-\frac{b^2+b_0^2}{2\sigma^2}} \nonumber \, ,
\end{eqnarray}
with $I_{0}$ being the modified Bessel function of the first kind \cite{AbS72}. Similar to before, the partial multipole rates in Eq.~(\ref{eq_transition_probability_twisted_wave_delocalized}) are given by the product of their plane--wave counterparts $W_{fi}^{\rm (pl)}(p L)$ and the function that depends on the properties of twisted beam and geometry of atomic target.

\subsubsection{Relative contributions of multipole transitions}
\label{subsubsec:relative_contribution_twisted}

As mentioned already above, the summation over multipole components ($pL$) in  Eqs.~(\ref{eq_transition_probability_twisted_wave}) and (\ref{eq_transition_probability_twisted_wave_delocalized}) is restricted by the parity and angular momentum
selection rules. Rather often, therefore, an atomic transition $\ketm{\alpha_i J_i} + \gamma \to \ketm{\alpha_f J_f}$ may proceed via just two channels and its rate is given then by:
\begin{equation}
	\label{eq:twisted_two_channels_probability}
	W_{fi}^{\rm (tw)} = W_{fi}^{\rm (tw)}(p_1 L_1) + W_{fi}^{\rm (tw)}(p_2 L_2) \, .
\end{equation}
In the standard plane--wave case the contribution of these channels is defined solely by the electronic structure of the target atom. This is not the case, however, for twisted light for which the partial multipole rates  $W_{fi}^{\rm (tw)}(p L)$ depend also on the geometrical properties of the radiation and the position of the atom. In order to better understand how the (relative) strength of multipoles is affected by the interaction with twisted light we consider the probability ratio:
\begin{equation}
	\label{eq:twisted_well_localized_ratio}
	\frac{W_{fi}^{\rm (tw)}(p_1 L_1)}{W_{fi}^{\rm (tw)}(p_2 L_2)} = \mathcal{R}_{L_1 L_2}\left(\varkappa b\right) \, \, 
	\frac{W_{fi}^{\rm (pl)}(p_1 L_1)}{W_{fi}^{\rm (pl)}(p_2 L_2)} \, , 
\end{equation}
which is written as a product of two terms. The first term is the geometrical factor
\begin{equation}
	\label{eq:R_factor_well_localized}
	\mathcal{R}_{L_1 L_2}\left(\varkappa b\right) = \frac{\sum\limits_{M} \left| d^{L_1}_{M \lambda}(\theta_k) \, J_{m-M}(b\varkappa) \right|^2}{\sum\limits_{M} \left| d^{L_2}_{M \lambda}(\theta_k) \, J_{m-M}(b\varkappa) \right|^2}
\end{equation}
that depends on the TAM projection $m$ and the opening angle $\theta_k$ of the Bessel beam as well as on the position of the target atom. The second term in Eq.~(\ref{eq:twisted_well_localized_ratio}) is just a squared ratio of the reduced matrix elements of multipole transitions:
\begin{equation}
	\label{eq:plane_wave_probabilities_ratio}
	\frac{W_{fi}^{\rm (pl)}(p_1 L_1)}{W_{fi}^{\rm (pl)}(p_2 L_2)} = \frac{\left|\rmem{\alpha_f J_f}{H_\gamma(p_1 L_1)}{\alpha_i J_i} \right|^2}{\left|\rmem{\alpha_f J_f}{H_\gamma(p_2 L_2)}{\alpha_i J_i} \right|^2}\, . 
\end{equation}
This expression is independent of the geometry of the process and just reflects the electronic structure of the atom. 

Eq.~(\ref{eq:twisted_well_localized_ratio}) clearly indicates that the relative strength of multipole transitions, if induced by twisted light, is modified with regard to standard plane-wave radiation. For an atom with well--defined impact parameter $b$, this modification is determined by the factor $\mathcal{R}_{L_1 L_2}\left(\varkappa b\right)$.
To consider a more realistic scenario of delocalized atom we need to use the \textit{averaged} geometrical factor:
\begin{equation}
	\label{eq:R_factor_delocalized}
	{\tilde \mathcal{R}}_{L_1 L_2}\left(\varkappa b_0 \right) = \frac{\sum\limits_{M} \left| d^{L_1}_{M \lambda}(\theta_k) \right|^2 {\mathcal J}_{m-M}(b_0) }{\sum\limits_{M} \left| d^{L_2}_{M \lambda}(\theta_k) \right|^2 {\mathcal J}_{m-M}(b_0)} \, ,
\end{equation}
where ${\mathcal J}_{m-M}(b_0)$ is given by Eq.~(\ref{eq_cal_J_function}). Being inserted into Eq.~(\ref{eq:twisted_well_localized_ratio}) in place of $\mathcal{R}_{L_1 L_2}\left(\varkappa b\right)$ this averaged factor can help to analyze the role of different multipole transitions in atomic trap experiments.  

\begin{figure}
	\centering
	\includegraphics[width=0.9\linewidth]{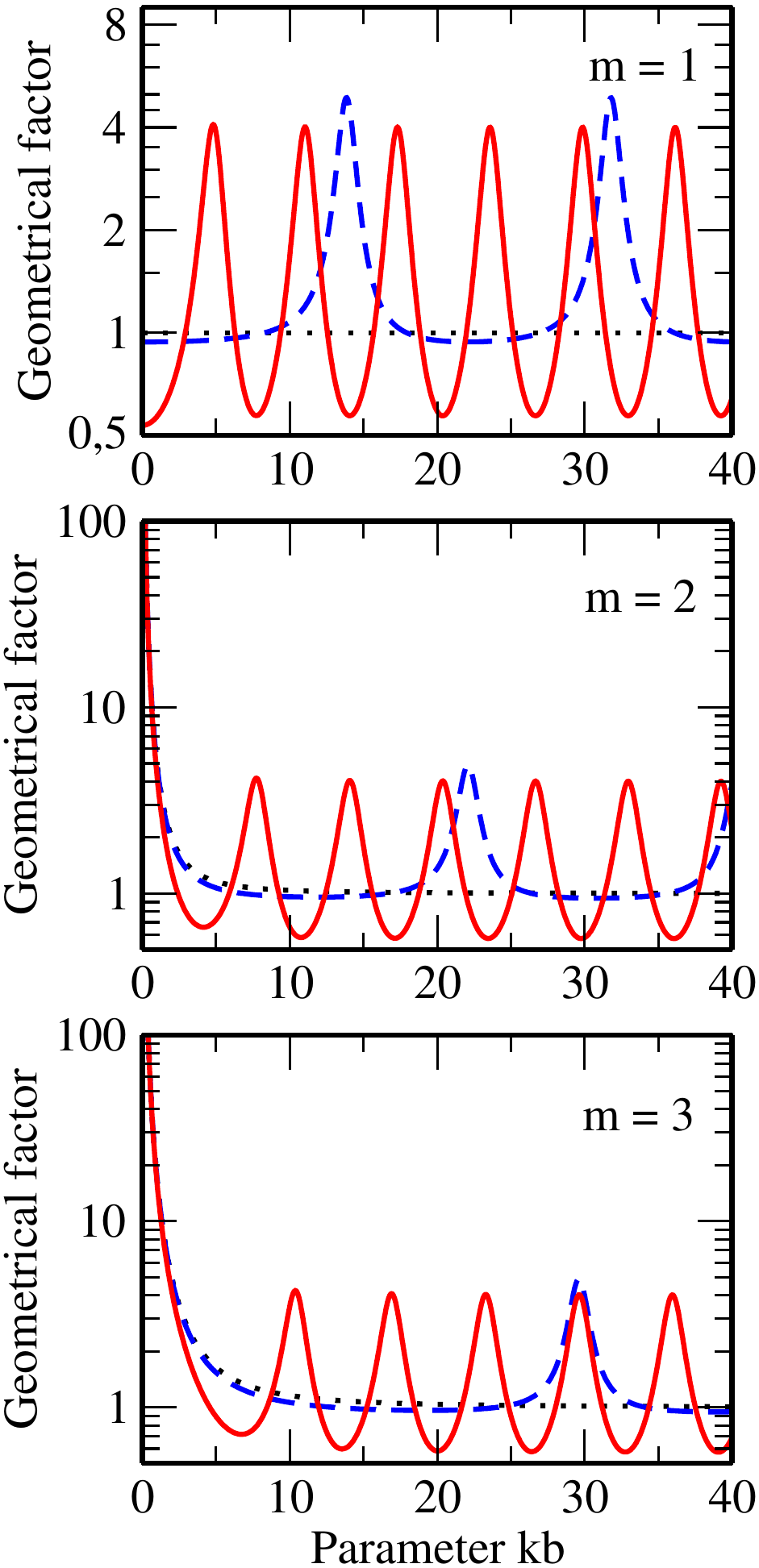}
	\vspace*{0cm}
	\caption{(Color online) The geometrical factor $\mathcal{R}_{L_1 = 2, \, L_2 = 1}\left(\varkappa b\right)$ as relevant for all $M2/E1$ and $E2/M1$ probability ratios. Calculations have been performed for the incident Bessel light with helicity $\lambda = 1$,  TAM projections $m = 1$ (upper panel), $m = 2$ (middle panel) and $m = 3$ (lower panel) and with opening angles $\theta_k = 1$~deg (black dotted line), $\theta_k = 10$~deg (blue dashed line) and $\theta_k = 30$~deg (red solid line).}
	\label{Fig1}
\end{figure}
%
%

%
%
\section{Results and discussion}
\label{sec:results_discussion}

\subsection{Geometrical factors for Bessel light}
\label{subsec:results_geometrical_factors}

In the previous section we have shown that the ratios of multipole transition probabilities for plane-- and twisted waves are just related to each other by the geometrical factors (\ref{eq:R_factor_well_localized}) and (\ref{eq:R_factor_delocalized}). The \textit{magnitude} of these factors determines how strongly the twisted light affects the relative contribution of multipole terms. Hence, it is instructive to calculate $\mathcal{R}_{L_1 L_2}\left(\varkappa b\right)$ and ${\tilde \mathcal{R}}_{L_1 L_2}\left(\varkappa b_0 \right)$ for some typical pairs of multipoles as they appear in atomic spectroscopy. In Fig.~\ref{Fig1}, for example, we display the geometrical factor $\mathcal{R}_{L_1 = 2, \, L_2 = 1}\left(\varkappa b\right)$ that describes the enhancement (or reduction) of the $M2/E1$ or $E2/M1$ probability ratios. This factor is calculated as a function of the dimensionless parameter $k b = \alpha \, \omega \, b$ and for three TAM projections, $m = 1$ (upper panel), $m = 2$ (middle panel) and $m = 3$ (lower panel). We have assumed, moreover, that the incident Bessel beam is a superposition of  plane--waves with helicity $\lambda = 1$ and with linear momenta ${\bm k}$ that lay on the surface of a cone with the opening angles $\theta_k = 1$~deg (dotted line), $\theta_k = 10$~deg (dashed line) and $\theta_k = 30$~deg (solid line), see Eq.~(\ref{eq_vector_potential_twisted}). As seen from the figure, the geometrical factor $\mathcal{R}_{L_1 = 2, \, L_2 = 1}\left(\varkappa b\right)$ is very sensitive to both, the opening angle $\theta_k$ and the TAM projection $m$. In particular, for a small opening angle $\theta_k = 1$~deg and for $m = 1$ the $\mathcal{R}_{L_1 L_2}\left(\varkappa b\right)$ is almost \textit{unity} thus indicating no change in the $M2/E1$ (or $E2/M1$) probability ratios in comparison to the plane--wave case. This is well expected since in the limit $\theta_k \to 0$ and for $m = \lambda$ the twisted--wave vector potential (\ref{eq_vector_potential_twisted})--(\ref{eq_vector_potential_twisted_expansion_coefficients}) recovers the standard solution for a plane wave propagating along the $z$--axis \cite{MaH13}. With the increase in $\theta_k$ the geometrical factor may significantly deviate from the plane--wave prediction $\mathcal{R}^{(pl)} = 1$. For $k b \gtrsim 5$, for example, the $\mathcal{R}_{L_1 = 2, \, L_2 = 1}\left(\varkappa b\right)$ oscillates and can reach the value of about 5 which implies significant enhancement of the quadrupole ($L = 2$) term with respect to the dipole ($L = 1$) counterpart. 

\begin{figure}
	\centering
	\includegraphics[width=0.9\linewidth]{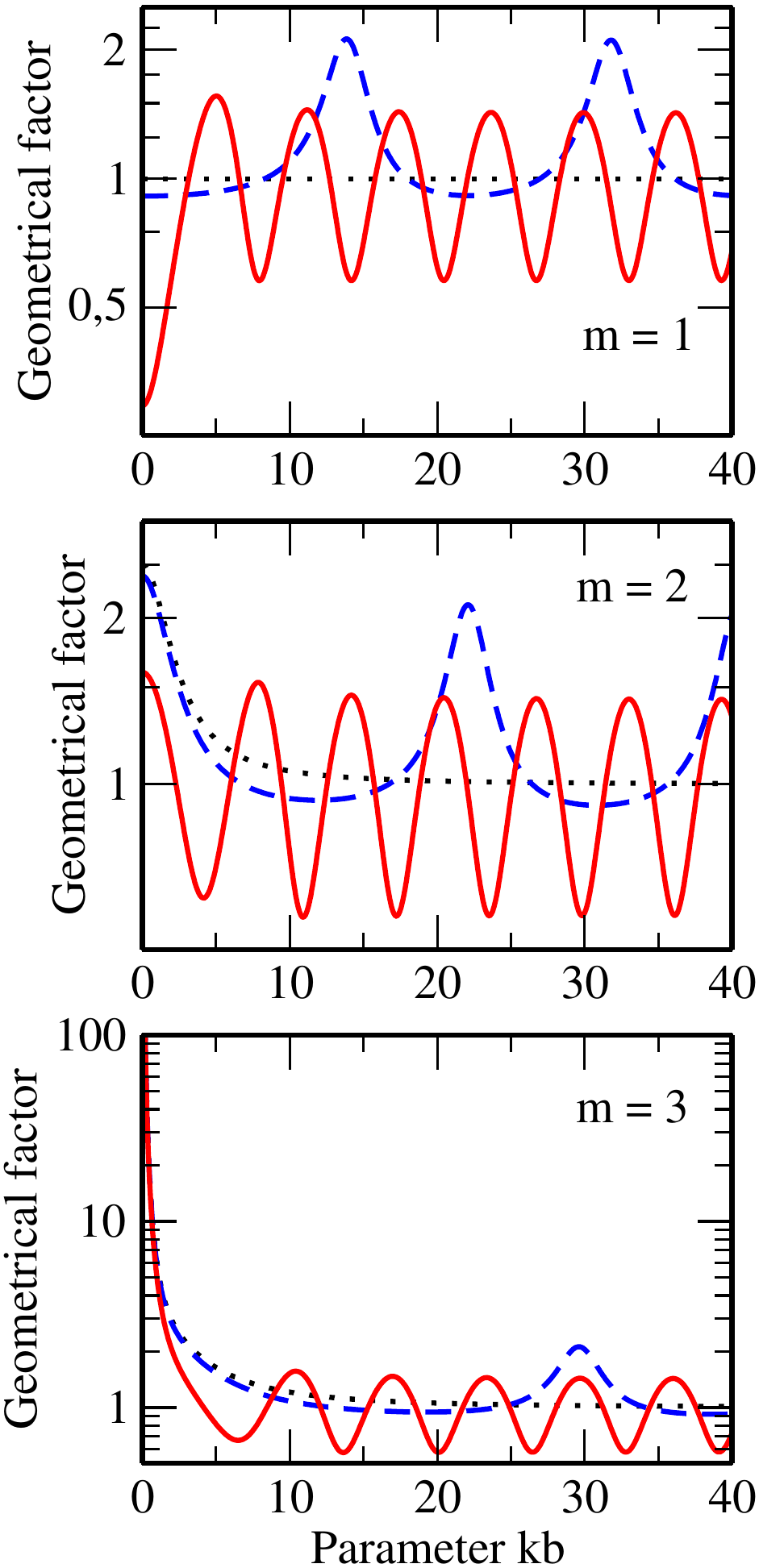}
	\vspace*{0cm}
	\caption{(Color online) The same as Fig.~\ref{Fig1} but for the geometrical factor $\mathcal{R}_{L_1 = 3, \, L_2 = 2}\left(\varkappa b\right)$ for the $E3/M2$ and $M3/E2$ probability ratios.}
	\label{Fig2}
\end{figure}

The interaction with Bessel light is most pronounced for the TAM projection $m > 1$ and small impact parameters, $k b \to 0$. In this parameter range the geometrical factor $\mathcal{R}_{L_1 = 2, \, L_2 = 1}\left(\varkappa b\right)$ is orders of magnitudes larger than $\mathcal{R}^{(pl)} = 1$. This enhancement is observed for all (non--zero) opening angles $\theta_k$ and can be explained based on the \textit{selection rules} for a transition $\ketm{\alpha_i J_i M_i} + \gamma \to \ketm{\alpha_f J_f M_f}$ induced by twisted light. Generally, these modified selection rules strongly depend on the position of the target atom in the Bessel wave front. For the particular case of $b \to 0$, when atom is located at the beam axis, the magnetic quantum numbers $M_i$ and $M_f$ of the initial and final state are related to the TAM projection $m$ of the Bessel beam by
\begin{equation}
	\label{eq:twisted_selection_rule}
	M_f = m + M_i \, . 
\end{equation}
From this expression and Eq.~(\ref{eq_transition_amplitude_twisted_wave_expansion}) it immediately follows that only multipoles with $L \geq \left|m\right|$ can contribute to the excitation. For $m = 2$ and small impact parameters, therefore, dipole transitions are strongly suppressed thus resulting in a drastic enhancement of $\mathcal{R}_{L_1 = 2, \, L_2 = 1}\left(\varkappa b\right)$. With the further increase in the TAM projection, \mbox{$m > 2$}, \textit{both} $L = 1$ and $L = 2$ terms are forbidden for $b = 0$ but the quadrupole transition rate grows much faster than the dipole one as an atom moves away from the beam center. Again, this leads to very large values of $\mathcal{R}_{L_1 = 2, \, L_2 = 1}\left(\varkappa b\right)$ for $k b \lesssim 1$.

Until now we have considered the geometrical factor $\mathcal{R}_{L_1 = 2, \, L_2 = 1}\left(\varkappa b\right)$ that characterizes the ratio of quadrupole to dipole transitions probabilities. In order to discuss how the absorption of twisted photons can affect other pairs of multipoles, we display in Fig.~\ref{Fig2} the factor $\mathcal{R}_{L_1 = 3, \, L_2 = 2}\left(\varkappa b\right)$. As follows from Eq.~(\ref{eq:twisted_well_localized_ratio}), this factor describes the $M2/E3$ or $E2/M3$ probability (or rate) ratios. Again, calculations have been performed for three TAM projections, $m = 1, 2$, and 3, and three opening angles, $\theta_k$ = 1 deg, 10 deg, and 30 deg, of the incident Bessel beam. Similar to the quarupole--to--dipole case, $\mathcal{R}_{L_1 = 3, \, L_2 = 2}\left(\varkappa b\right)$ is strongly dependent on these two parameters that characterize the twisted light. The $\theta_k$--dependence is very pronounced for an atom displaced from the beam center by $k b \gtrsim 5$. In this case the geometrical factor $\mathcal{R}_{L_1 = 3, \, L_2 = 2}\left(\varkappa b\right)$ reproduces almost identically the plane--wave result $\mathcal{R}^{(pl)} = 1$ for $\theta_k = 1$ deg while it oscillates for large opening angles $\theta_k = 10$ deg and $\theta_k = 30$ deg. Even though the amplitude of these oscillations is smaller than for the quadrupole--to--dipole case, it indicates that $M2/E3$ and $E2/M3$ rate ratios can be increased (or decreased) by almost a factor of two if one chooses a proper position of the target atom in the light beam. 

As in the $M2/E1$ (or $E2/M1$) case, c.f. Fig.~\ref{Fig1}, the largest enhancement of the factor  $\mathcal{R}_{L_1 = 3, \, L_2 = 2}\left(\varkappa b\right)$ can be found for very small impact parameters, $k b \to 0$. In contrast to $\mathcal{R}_{L_1 = 2, \, L_2 = 1}\left(\varkappa b\right)$, however, the octupole--to--quadrupole factor increases by orders of magnitude only for TAM projections $m > 2$. This is again a consequence of the transition selection rules which allow both quadrupole ($L = 2$) and octupole ($L = 3$) excitations of an atom, located at the center of the Bessel beam with $m \leq 2$. Only for the higher topological charge $m = 3$ the $M2$ (or $E2$) transition is suppressed which leads to a significant enhancement of the factor $\mathcal{R}_{L_1 = 3, \, L_2 = 2}\left(\varkappa b\right)$, see lower panel of Fig.~\ref{Fig2}.    

\begin{figure}
	\centering
	\includegraphics[width=0.9\linewidth]{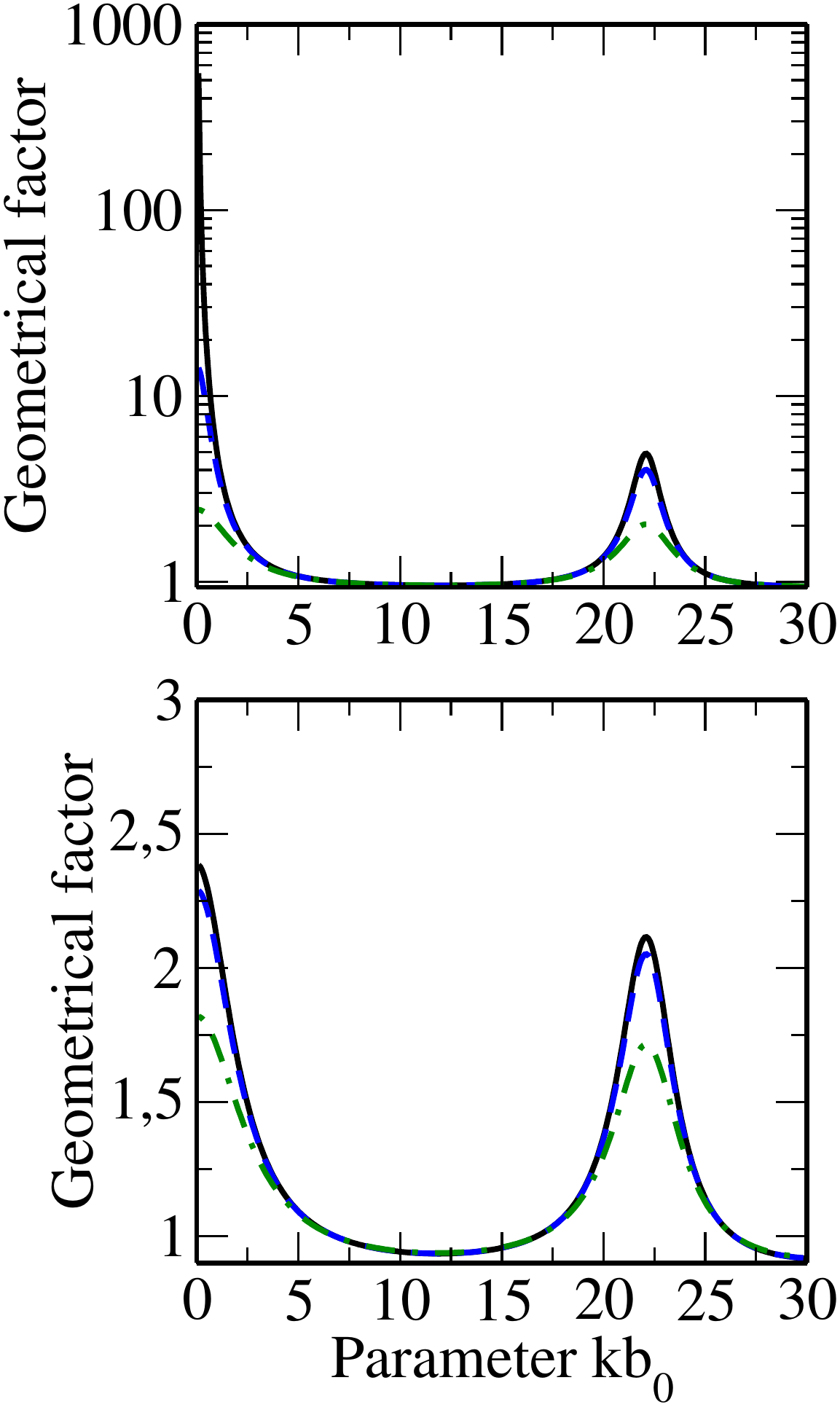}
	\vspace*{0cm}
	\caption{(Color online) The averaged geometrical factors ${\tilde \mathcal{R}}_{L_1 = 2, \, L_2 = 1}\left(\varkappa b_0 \right)$ (upper panel) and ${\tilde \mathcal{R}}_{L_1 = 3, \, L_2 = 2}\left(\varkappa b_0 \right)$ (lower panel). Calculations have been performed for the incident Bessel light with helicity $\lambda = 1$,  TAM projection $m = 2$ and opening angle $\theta_k = 10$~deg. The position of a target atom is described by the Gaussian distribution (\ref{eq_atom_localization_probability}) with the width $\sigma = 1 /\left(15 k \right)$ (blue dashed line) and $\sigma = 1/\left(5 k \right)$ (green dashed--dotted line). These results are compared, moreover, with the prediction obtained for a well--localized atom (black solid line).}
	\label{Fig3}
\end{figure}

Figs.~\ref{Fig1} and \ref{Fig2} display the quadrupole--to-dipole and octupole--to--quadrupole factors $\mathcal{R}_{L_1, \, L_2}\left(\varkappa b\right)$ calculated for a target atom with well--defined position with respect to the center of the Bessel beam. As we mentioned already, this idealized case can not be realized in the present experiments where the impact parameter $b$ is determined with a rather large uncertainty. In order to investigate the effect of this uncertainty for the relative strength of multipole transitions we compare in Fig.~\ref{Fig3} the geometrical factors for a well--localized (solid line) and delocalized atom. For the latter case we employed Eq.~(\ref{eq:R_factor_delocalized}) with a width of the position distribution $\sigma = 1 /\left(15 k \right)$ (dashed line) and $\sigma = 1/\left(5 k \right)$ (dashed--dotted line). Again, the quadrupole--to-dipole and octupole--to--quadrupole factors have been evaluated as functions of the dimensionless parameter $k b_0$, where $b_0 = b$ for the ``localized--atom'' case. In contrast to Figs.~\ref{Fig1} and \ref{Fig2}, however, we have restricted here our calculations to a single TAM projection $m = 2$ and the opening angle $\theta_k = 10$~deg. As seen from the figure, the delocalization of the target atom causes the geometrical factor ${\tilde \mathcal{R}}_{L_1 L_2}\left(\varkappa b_0 \right)$ to approach the plane--wave result $\mathcal{R}^{(pl)} = 1$. The most pronounced effect can be observed for the quadrupole--to-dipole factor and small impact parameters, $b \to 0$. For this case  ${\tilde \mathcal{R}}_{L_1=2, L_2=1}\left(\varkappa b_0 \right)$ is reduced by at least one order of magnitude if the position of the atom within the Bessel wave--front is not well defined. Also at $k b_0 \approx 22$ the geometrical factor ${\tilde \mathcal{R}}_{L_1=2, L_2=1}\left(\varkappa b_0 \right)$ decreases by almost factor three with the increase in the location uncertainty $\sigma$. These results clearly indicate the demand for a precise localization of the target atom for an efficient modification of the relative strength of multipole atomic transitions by a twisted light.

\subsection{$^3P_1 \to ^1P_1$ transition in neutral Mg}
\label{subsec:results_Mg_transition}

The geometrical factors $\mathcal{R}_{L_1 \, L_2}\left(\varkappa b\right)$ and ${\tilde \mathcal{R}}_{L_1 L_2}\left(\varkappa b_0 \right)$, discussed in the previous section, allow one to analyze the relative contribution of multipole terms for an \textit{arbitrary} atom and \textit{arbitrary} transition. Of special interest here is the analysis of electric--dipole forbidden excitations. In atomic spectroscopy a number of such transitions is known which proceed via several \textit{non}--E1 channels of almost \textit{equal} (plane--wave) strength. In this case the effect of absorption of twisted photons will be very pronounced and can be easily observed experimentally. In contrast, for transitions involving electric dipole term the plane--wave $pL/E1$ probability ratios are usually very small. It will be therefore very difficult to approve changes in the relative contributions of $pL$ and $E1$ multipole components if twisted light interacts with delocalized atom.

As an example of electric--dipole forbidden transition we consider here the $3s3p \, {}^3P_1 \to 3s3p \, {}^1P_1$ excitation in neutral Mg. This transition, whose wavelength is 758 nm and which
is well separated from—much stronger—electric dipole transitions \cite{Kra19}, can be induced by the standard red diode lasers.
It may proceed, moreover, via either magnetic dipole M1
or electric quadrupole E2 channels. In order to estimate the
probabilities of these multipole transitions, we have used
the multiconfiguration Dirac-Fock (MCDF) approach which
allows to account for the relativistic and magnetic interaction
effects in many-electron systems. The detailed MCDF calculations have been performed with the help of the recently
developed JAC code \cite{Fri19} and have revealed the oscillator
strengths $f_{M1} = 3.5 \times 10^{-12}$ and $f_{E2} = 4.8 \times 10^{-12}$  for the
M1 and E2 transitions, respectively.
Hence, if the ${}^3P_1 \to {}^1P_1$ transition is induced by plane--wave radiation, the rates of both channels will be almost the same, $W_{fi}^{\rm (pl)}(E2)/W_{fi}^{\rm (pl)}(M1) \approx 1.37$. As discussed above, one can use Bessel light in order to modify this ratio. For example, the contribution of the quadrupole transition can be significantly enhanced if the Mg atom is placed near the center of the Bessel beam with TAM projection $m = 2$, see middle panel of Fig.~\ref{Fig1}. The size of this enhancement depends, however, on the uncertainty $\sigma$ of the target position. For example, as seen from Fig.~\ref{Fig3} an order--of--magnitude increase in the averaged geometrical factor ${\tilde \mathcal{R}}_{L_1=2, L_2=1}\left(\varkappa b_0 = 0 \right)$ and, hence, of the ratio $W_{fi}^{\rm (tw)}(E2)/W_{fi}^{\rm (tw)}(M1)$, can be observed if $\sigma = 1/(5k)$. For the ${}^3P_1 \to {}^1P_1$ transition with energy $\hbar \omega = 1.84$~eV this corresponds to $\sigma = 21$~nm; the position uncertainty which is comparable to that achieved in present experiments \cite{ScS16}. 

\begin{figure}
	\centering
	\includegraphics[width=0.9\linewidth]{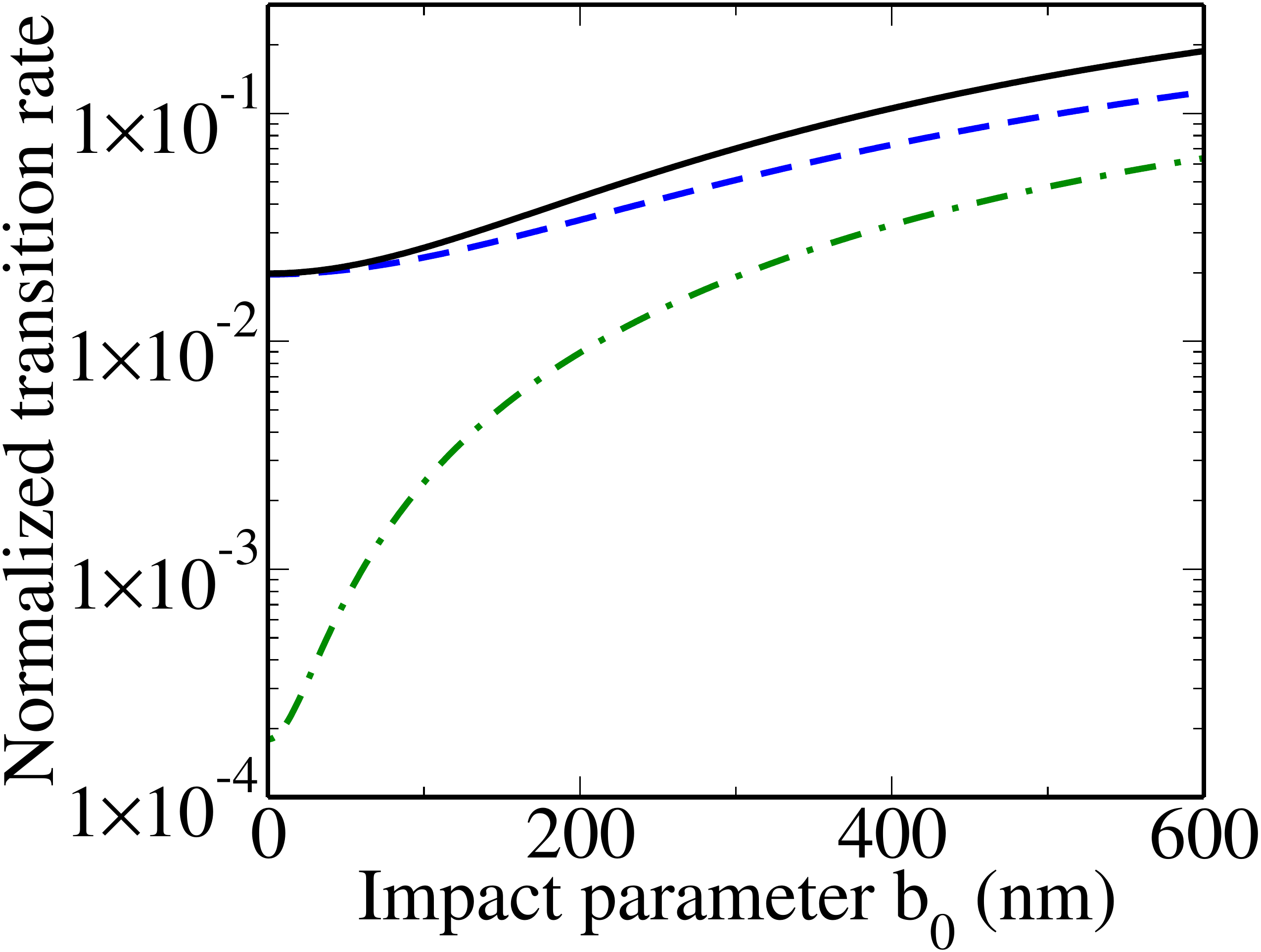}
	\vspace*{0cm}
\caption{
	(Color online) The rates of the magnetic dipole (green dashed--dotted line) and electric quadrupole (blue dashed line) components of the $3s3p \, {}^3P_1 \to 3s3p \, {}^1P_1$ photo--excitation of neutral Mg atom. The sum of both terms, $W^{\rm (tw)}_{fi} = W^{\rm (tw)}_{fi}(M1) + W^{\rm (tw)}_{fi}(E2)$, is displayed by the black solid line.
	Calculations have been performed for the incident Bessel light with helicity $\lambda = 1$,  TAM projection $m = 2$ and opening angle $\theta_k = 10$~deg. The target Mg atom is placed at the impact parameter $b_0$ with the uncertainty $\sigma$~=~21~nm. The rates are normalized with respect to the total rate $W^{\rm (pl)}_{fi}$ of the excitation by plane--wave light.  
	}
	\label{Fig4}
\end{figure}

The magnetic dipole (dashed--dotted line) and electric quadrupole (dashed line) transition rates, obtained for the uncertainty $\sigma =$21~nm from Eq.~(\ref{eq_transition_probability_twisted_wave_delocalized}), are displayed in Fig.~\ref{Fig4} as functions of the impact parameter $b_0$. These rates, as well as their sum (solid line), are normalized with respect to the total plane--wave counterpart $W^{\rm (pl)}_{if}$ and are computed for an incident Bessel beam with opening angle $\theta_k =$~10~deg and TAM projection $m =$~2. As seen from the figure, $W_{fi}^{\rm (tw)}(M1)$ is strongly suppressed near the beam center as it can be understood from the modified selection rule  (\ref{eq:twisted_selection_rule}). For $b_0 = 0$, therefore, the ${}^3P_1 \to {}^1P_1$ transition proceeds predominantly via the $E2$ channel; the effect that has been expected from the analysis of the averaged geometrical factor  ${\tilde \mathcal{R}}_{L_1=2, L_2=1}\left(\varkappa b_0 \right)$.

As seen from Fig. \ref{Fig4}, the rate for the excitation induced
by the Bessel light is smaller comparing to its plane-wave
counterpartner; the effect which is mostly pronounced for
small $b_{0}$. This can be well expected since for the rather small
spread, $\sigma = 21$ nm of the impact parameter the target atom is
exposed to low light intensity near the beam axis. Of course,
the twisted excitation rate can be significantly increased by
enhancing the $\sigma $ and can even become comparable to the $W^{(pl)}_{fi}$
if $\sigma \rightarrow \infty$. In this case, however, no modification of the
multipole transitions by twisted light will be possible since
 ${\tilde \mathcal{R}}_{L_1=2, L_2=1}\left(\varkappa b_0 \right) \approx 1$, see Fig. \ref{Fig3} and Eq. (\ref{eq:R_factor_delocalized}). Therefore, the
precise position control of the trapped atom is necessary to
observe the twisted light effect.

%
%
\section{Summary and outlook}
\label{sec:summary}

In summary, we have performed a theoretical study of the excitation of a single trapped
atom or ion by incident light. Special attention was paid to the case when the photo--excitation proceeds via several (electric and magnetic) multipole channels. We have shown that the relative strength of these channels can be significantly modified if instead of the standard plane--wave radiation one uses \textit{twisted} Bessel light. In order to quantify the effect of the ``twistedness'' on the (ratio of) multipole transition rates we have introduced the geometrical factors (\ref{eq:R_factor_well_localized}) and (\ref{eq:R_factor_delocalized}). These factors are   independent of the electronic structure of a particular atom but reflect the kinematic properties and topological charge of the incident Bessel beam as well as position of a target atom in such a beam. Moreover, the \textit{averaged} geometrical factor ${\tilde \mathcal{R}}_{L_1, L_2}\left(\varkappa b_0 \right)$ accounts for realistic scenario in which the position of an atom is not well determined. The use of this factor may help in the preparation and analysis of experiments with single trapped ions. 

Although the geometrical factors ${\mathcal{R}}_{L_1, L_2}\left(\varkappa b \right)$ and ${\tilde \mathcal{R}}_{L_1, L_2}\left(\varkappa b_0 \right)$ are \textit{universal} and can be used for \textit{any} atomic system, we have applied them here to investigate the $3s3p \, ^3P_1 \to 3s 3p \, ^1P_1$ excitation of a neutral Mg atom. This transition can proceed via either magnetic dipole (M1) or electric quadrupole (E2) channels whose probabilities are almost the same if the atom is exposed to  plane--wave radiation. In contrast, the application of the Bessel light can enhance the relative strength of the E2 transition by almost an order of magnitude. Based on the analysis of the averaged geometrical factor ${\tilde \mathcal{R}}_{L_1, L_2}\left(\varkappa b_0 \right)$ we have shown that this enhancement can be achieved if the Mg atom is placed at the beam axis with the maximal uncertainty $\sigma \approx$~20 nm.

Although our present computations were carried out for a  Bessel beam only, the results obtained here are also applicable to the excitation with twisted Laguerre-Gaussian beams if the target atom is placed close to the beam center. This is due to the fact that both the paraxial Bessel and Laguerre-Gaussian beams with non-zero OAM $m_{l}$ behave like $r^{m_{l}}e^{im_{l}\phi}$ for small $r$.

In the present study we have focused on transitions between fine--structure levels $\ketm{\alpha J}$, whose magnetic substates remains unobserved. However, the photo--excitation of trapped atoms prepared in a particular substate $\ketm{\alpha J M_J}$ is also a subject of considerable experimental and theoretical interest. The multipole components of such $\ketm{\alpha_i J_i M_{i}} + \gamma \to \ketm{\alpha_f J_f M_{f}}$ transitions can be strongly affected by twisted light as well. Apart of the ``usual'' enchancement (or reduction) of transition rates $W_{fi}^{\rm (tw)}(pL)$, one might also expect the modification of the \textit{multipole--mixing} terms that arise in magnetic--sublevel transitions. Detailed analysis of these effects of twisted beams is currently underway and will be presented soon.

\section*{ACKNOWLEDGEMENTS}
We gratefully acknowledge the support of the Braunschweig International Graduate School of Metrology B-IGSM and the DFG Research Training Group 1952 Metrology for Complex Nanosystems.

%
%
%
%


\begin{thebibliography}{12}
	
\bibitem{BeB11} A.~Bekshaev, K.~Y.~Bliokh, and M.~Soskin, 
				J. Opt. {\bf 13}, 053001 (2011).

\bibitem{MaH13} O.~Matula, A.~G.~Hayrapetyan, V.~G.~Serbo, A.~Surzhykov, and S.~Fritzsche, 
				J. Phys. B {\bf 46}, 205002 (2013).

\bibitem{SuS16} A.~Surzhykov, D.~Seipt, and S.~Fritzsche,
                Phys. Rev. A {\bf 94}, 033420 (2016). 

\bibitem{PeV18} A.~A.~Peshkov, A.~V.~Volotka, A.~Surzhykov, and S.~Fritzsche,
				Phys. Rev. A {\bf 97}, 023802 (2018).

\bibitem{JeS11} U.~D.~Jentschura and V.~G.~Serbo,
				Phys. Rev. Lett. {\bf 106}, 013001 (2011).

\bibitem{Vie16} J.~Vieira \textit{et al.}, 
				Nature communications {\bf 7}, 10371 (2016).

\bibitem{ScS16} C.~T.~Schmiegelow, J.~Schulz, H.~Kaufmann, T.~Ruster, U.~G.~Poschinger, and F.~Schmidt-Kaler,   
				Nat. Commun. {\bf 7}, 12998 (2016).

\bibitem{AfC13} A.~Afanasev, C.~E.~Carlson, and A.~Mukherjee,
                Phys. Rev. A {\bf 88}, 033841 (2013).
                
\bibitem{AfC18} A.~Afanasev, C.~E.~Carlson, C.~T.~Schmiegelow, J.~Schulz, F.~Schmidt-Kaler, and M.~Solyanik,
                New J. Phys. {\bf 20}, 023032 (2018).

\bibitem{AfC18b} A. Afanasev, C.~E.~Carlson, and M.~Solyanik,	
                Phys. Rev. A {\bf 97}, 023422 (2018). 

\bibitem{Ros57} M.~E.~Rose,
				{\it Elementary theory of angular momentum} (John Wiley \& Sons, New York, 1957).

\bibitem{Gra74} I.~P.~Grant,
				J. Phys. B {\bf 7}, 1458 (1974).

\bibitem{SuF02} A.~Surzhykov, S.~Fritzsche, and Th.~St\"ohlker,
				J. Phys. B {\bf 35} (2002) 3713.

\bibitem{BrJ83} B.~H.~Bransden, C.~J.~Joachain, {\it Physics of atoms and molecules} (Longman, New York, 1983).

\bibitem{ScF14} H.~M.~Scholz-Marggraf, S. Fritzsche, V. G. Serbo, A. Afanasev, and A. Surzhykov, Phys. Rev. A {\bf 90}, 013425 (2014).

\bibitem{AbS72} M.~Abramowitz, I.~A.~Stegun, {\it Handbook of Mathematical Functions} (Dover, New York, 1972).

\bibitem{Kra19} A.~Kramida, Yu.~Ralchenko, J.~Reader, and NIST ASD Team,
NIST Atomic Spectra Database, version 5.6.1 (National Institute of Standards and Technology, Gaithersburg, MD, 2019),
available: https://physics.nist.gov/asd (18 July 2019).

\bibitem{Fri19} S.~Fritzsche, Comput. Phys. Commun. 240 (2019) 1-14.
\end{thebibliography}
\end{document}